\documentstyle[pra,aps,preprint]{revtex}
\begin{document}
\draft
\title{Role of Orbitals in Manganese Oxides  - Ordering and Fluctuation}
\author{R. Maezono, S. Murakami, N. Nagaosa}
\address{Department of Applied Physics, University of Tokyo,
Bunkyo-ku, Tokyo 113-8656, Japan}
\author{S. Ishihara}
\address{Institute for Materials Research, Tohoku University, 
Sendai, 980-8677, Japan.}
\author{M. Yamanaka}
\address{
Department of Applied Physics, Science University of Tokyo,
Kagurazaka, Shinjuku-ku, Tokyo 162, Japan.}
\author{H.C. Lee}
\address{
Asia Pacific Center for Theoretical Physics, Seoul 151-742,
Korea.}
\maketitle
\begin{abstract}
We study the manganese oxides from the viewpoint of the 
strongly correlated doped Mott insulator.
The magnetic ordering and the charge transport are governed by the orbital
degrees of freedom, and their dimensionality is controlled by the
anisotropic transfer integrals between the $e_g$ orbitals.
As $x$ increases the  magnetic structure is predicted to 
change as  $A \to F \to A \to C \to G$
( $F$: ferromagnet, $A$: layered antiferromagnet, 
$C$: rod-type antiferromagnet, $G$: usual antiferromagnet),
in agreement with experiments. Especially the orbital 
is aligned as $d_{x^2-y^2}$ in the metallic 
$A$ state, which explains the quasi 2D transport and no canting 
of the spin observed experimentally.

Next we discuss the ferromagnetic state without the orbital ordering due to 
the quantum fluctuation. Here the interplay between the electron repulsion 
$U$ and the Jahn-Teller
electron-phonon interation $E_{\rm LR}$ is studied with a large $d$ model. 
In addition to this strong correlation, we propose that the dynamical phase 
separation could explain the specific heat as well as the various
anomalous physical properties, e.g., resistivity, photo-emission,
etc.
\end{abstract}
\narrowtext
\section{Introduction}
The interplay between the magnetism and transport properties is one of the 
main issues in the physics of the strongly correlated electronic systems.
Doped manganites R$_{1-x}$A$_x$MnO$_3$ (R=La, Pr, Nd, Sm ;
A= Ca, Sr, Ba) have recently attracted considerable interests 
from this viewpoint in connection to the colossal magnetoresistance (CMR).
It has been believed that the most fundamental mechanism for 
this CMR is the double exchange interaction, which  
has been studied long time ago \cite{Zener,Anderson,deGenne}.
However recent systematic experimental studies have revealed 
the vital role of the orbital degrees of freedom, which is the
subject of this paper.
\par
In the parent compound RMnO$_3$, the valence of the Mn ion is
Mn$^{3+}$ with the configuration $d^4$. 
Due to the crystal field the five $d$-orbitals 
split into two $e_g$-orbitals $(x^2-y^2, 3z^2-r^2)$
and three $t_{2g}$-orbitals $(xy, yz, zx)$.  
According to Hund's rule the four electrons occupy three 
$t_{2g}$-orbitals and one of the $e_g$-orbitals with parallel spins.
Then the $e_g$-electron has the orbital degrees of freedom in addition to
the spin ones.
When the holes with concentration $x$ are doped, the vacancies in the 
$e_g$-orbitals are introduced which induces the ferromagnetism due 
to the double exchange interaction and the charge transport.
However there are several experimtental facts which cannot be explained in
the simple double exchange models.
One is the rich phase diagrams which is studied theoretically in Section 
II and the other is the anomalous features in the ferromagnetic metallic phase
described in Section III.
All these issues are colsely related to the orbital degrees of freedom, 
which control the dimensioanlity of the electronic systems and 
whose quantum fluctuation gives rise to the orbital liquid and orbital
Kondo effect. 
Section IV is devoted to the conclusions 
where the possibility and implications of the dynamical phase
separation are mentioned.
\section{ Phase Diagram}
We set up the three-dimensional cubic lattice consisting of the 
manganese ions. 
Two kinds of the $e_g$ orbital( $\gamma, \gamma'$) 
are introduced in each site, 
and the $t_{2g}$ electrons are treated as the localized spin with $S=3/2$.
The Hamiltonian without the JT coupling is given as follows 
\cite{ishi96,ishi97,maezono1,maezono2}, 
\begin{eqnarray}
H
&=&\sum_{<ij> , \sigma , \gamma , \gamma'}  
\Bigl( t^{\gamma \gamma'}_{ij} 
d_{i \gamma \sigma}^\dagger  
d_{j \gamma' \sigma}^{\phantom{\dagger}}  
+h.c. \Bigr) \nonumber \\
&+&
U\sum_{i \gamma} 
n_{i \gamma \uparrow} n_{i \gamma \downarrow} 
+ U'\sum_{i}  n_{i a} n_{i b} 
+ I \sum_{i,  \sigma , \sigma'} 
d_{i a \sigma}^\dagger  d_{i b \sigma'}^\dagger
d_{i a \sigma'}^{\phantom{\dagger}}   
d_{i b \sigma }^{\phantom{\dagger}}    \nonumber \\
&+& J_H \sum_{i} 
{\vec S_i} \cdot {\vec S^{t_{2g}}_i} + J_s \sum_{<ij>} 
{\vec S^{t_{2g}}_i} \cdot {\vec S^{t_{2g}}_j} 
\ . 
\label{eq:ham}
\end{eqnarray}
$d_{i \gamma \sigma}^\dagger $ is the operator which 
creates an electron with spin $\sigma(=\uparrow, \downarrow)$ 
in orbital $\gamma(=a,b)$ at site $i$, and 
${\vec S_i}$ is the spin operator for the $e_g$ electron 
defined by 
${\vec S_i}=\frac{1}{2}\sum_{\sigma \sigma' \gamma} 
d^\dagger_{i \sigma \gamma} \vec \sigma_{\sigma \sigma'}
d_{i \sigma' \gamma} $.  
The first term describes the electron transfer 
between $\gamma$ orbital at site $i$ and 
$\gamma'$ orbital at the nearest-neighbor site $j$.  
The value of $t^{\gamma \gamma'}_{ij}$ 
is estimated by considering the oxygen $2p$ orbitals 
between the nearest Mn-Mn pair, and 
is represented by $c^{\gamma \gamma'}_{ij} t_0$, where 
$c^{\gamma \gamma'}_{ij}$ is the numerical factor depending 
on the orbitals \cite{ishi97}. 
$t_0$ is estimated to be $0.72 eV$, which we choose the unit of energy
below ($t_0=1$) \cite{ishi96}.
The second line shows the electron-electron interaction terms where
$U$, $U'$ and $I$
is the intra-, inter-orbital Coulomb interactions, and inter-orbital 
exchange interaction, respectively. 
This interaction can be rewritten as 
$
-\alpha \sum_i \Bigl(\vec S_i (\tau)
                      +\frac{J_H}{2 \alpha} \vec S_i^{t_{2g}}(\tau) \Bigr)^2
  -\beta \sum_i \vec T_i (\tau)^2                 
$ \cite{maezono2}.
Here the spin operator $\vec S_i$ and the iso-spin operator 
${\vec T_i}=\frac{1}{2}\sum_{\gamma \gamma' \sigma} 
d^\dagger_{i \sigma \gamma} \vec \sigma_{\gamma \gamma'}
d_{i \sigma \gamma'} $ for the orbital degrees of freedom
are introduced, and the two positive coefficients $\alpha$ and $\beta$, 
which are defined by 
$\alpha=2U/3+U'/3-I/6$ and $\beta=U'-I/2$ , represent the 
interaction to induce the spin and iso-spin moments, respectively.
The last line is the sum of the Hund coupling 
and the AF interaction between the nearest neighboring 
$t_{2g}$ spins. 
Here we adopt the mean field approximation
by introducing the order parameters 
$ \langle \vec S_i \rangle $,
$ \langle \vec S^{t_{2g}}_i \rangle $,
and 
$ \langle \vec T_i \rangle$.
These order parameters are determined to optimize the mean field energy
at zero temperature.
For both spin and orbital, the four types of the ordering are considered, 
that is, the ferromagnetic (F-type) ordering, where the order parameters 
are uniform, and the three AF-like  orderings, i.e., 
the layer-type (A-type), the rod-type (C-type) and the 
NaCl-type (G-type) AF orderings. 
Hereafter, types of the orderings are termed as, for example,
(spin:C), and so on. 
\par
In Fig. 1, we present the calculated phase diagram in the case 
of $\alpha = 8.1$, $\beta=2.5$, which is relevant to the 
actual manganese oxides. 
In each region, the spin structure is specified by the letter
F,A,C,G and the orbital configuration by picture.
The important results are the followings.

\noindent
[1] There occurs the crossover from the superexchange dominated region 
    for small $x$ ($x<0.3$) to that of the double exchange interaction 
    for large $x$ ($x>0.3$). The former region could not be reproduced
    by the analysis without Coulombic repulsion.
    In this respect, the spin F state up to $x\sim0.4$ in the phase diagram 
    is mostly due to the superexchange interaction.

\noindent
[2] Because of the anisotropy of the transfer integral between the 
    $e_g$-orbitals, the most natural orbital configuration expected from 
    the double exchange interaction is $(x^2-y^2)$ with the two 
    dimensional electronic dispersion. 
    This corresponds to the A phase for $0.15<x<0.45$.
    In this case the interlayer
    transfer is forbidden, and the spin A-type sturucture is stabilized 
    by $J_s$. No spin canting is expected in this case.

\noindent
[3] Similarly the spin C structure is accompanied by the $3z^2-r^2$
    orbital which gives the 1D-like dispersion along the $c$-axis. 
    Hence the spin and orbital structure is closely correlated through
    the dimensionality of the electronic dispersion. 
    From this viewpoint, the orbitals have more degrees of freedom
    in the ferromagnetic state. This suggests the large orbital fluctuation 
    in the spin F state as discussed in the next section.

\vskip 0.3cm
We now discuss the comparison between the mean field phase diagram in Fig. 1 
and the experiments. 
At the moment, a value of $J_S$ cannot be estimated accurately, 
but there are two rough estimates.
One is from the N\'eel temperature $T_N=130K$ for CaMnO$_3$ $(x=1.0)$
\cite{Woll},
which suggest $J_S = T_N/7.5 \cong  1.7 meV \cong 0.0023 t_0$
in the mean field approximation.
The fluctuations lower $T_N$, and hence increase the 
estimate for $J_S$. 
Another estimate is obtained from the numerical calculations 
for LaMnO$_3$ $(x=0.0)$, which suggests
$J_S \cong 8 m eV \cong 0.011 t_0$ \cite{ishi96}.
Although $J_S$ might depend on $x$ in real materials, 
we tentatively fix $J_S$ to be $0.009t_0$ represented by 
the broken line in Fig. 1.
Then the spin structure changes as A $\to $ F $\to $ A$\to$ C $\to$ G, 
as $x$ increases, which is in good agreement with the experiments
\cite{Kuwahara1,PrSr,LaSr}.
Another implication to the experimental results is
about spin:A phase appearing around $x>0.5$. 
In $\rm Nd_{1-x}Sr_xMnO_3$,
the ferromagnetic metallic phase is realized up to about $x=0.48$ and 
the CE-type AF structure with the charge ordering tunes up 
\cite{Kawano2,Kuwahara1}. 
With further increasing of $x$, 
the metallic state with spin:A again appears at about $x=0.53$,  
and the large anisotropy in the electrical resistivity is observed 
in this phase. 
The similar metallic phase accompanied with spin:A is also reported in 
$\rm Pr_{1-x}Sr_xMnO_3 $ \cite{PrSr}. 
The spin canting is not observed experimentally 
\cite{Kawano2,Kuwahara1,PrSr} as predicted 
above in this Spin:A metal, in contrast to the Spin:A in the small $x$ region
which has been discussed by de Gennes \cite{deGenne}. 
\par
Summarizing this section we have studied the phase diagram of 
La$_{1-x}$Sr$_x$MnO$_3$
in the plane of $x$ ( hole concentration ) and 
$J_s$ ( AF exchange interaction between the $t_{2g}$
spins) in the mean field approximation. 
The global features can be understood in terms of the interplay between the 
superexchange interaction and the double exchange one which 
is considerably modified with taking the orbital degrees 
of freedom into account. 
The dimensionality of the energy band 
attributed to the orbital structure plays an 
essential role to determine the phase boundary. 
The orbital structure in the ferromagnetic phase 
is sensitive to changes of the carrier concentration and the 
energy parameters. 
This is because the spin structure does not forbid the kinetic energy gain
for any direction, and the orbitals have more freedom.
This enhances the fluctuation of the orbital degrees of freedom.
Actually the orbital fluctuation has been studied in the slave-fermion
formalism, and it has been found that the orbitals remain
a liquid down to the low temperature due to the 
fluctuation among the three directions of the $x^2-y^2$-type 
ordering \cite{ishi97}.
What quenches the entropy of the system is the quantum 
tunneling between the different orbital configurations, which could be 
regarded as the orbital-Kondo effect. This will be studied in the 
next section.
\section{Orbital Fluctuation}
Experimentally there are several anomalous features 
in the low temperature ferromagnetic phase below $T_c$.
\par
\noindent
[a] In the neutron scattering experiment 
no temperature dependent phonon modes have been observed   \cite{endoh}. 
The recent Raman scattering experiment also shows that 
the Jahn-Teller phonons ( especially their frequencies )
are temperature independent and 
insensitive to the ferromagnetic transition at $T_c$ \cite{yamamoto}.
\noindent
[b] The photo-emission spectra show a small discontinuity at
 the Fermi edge even at $T<<T_c$, which suggests some interactions 
 still remain strong there \cite{sarma}. 
\noindent
[c] The optical conductivity $\sigma(\omega)$ at $T<<T_c$ is composed of
two components, i.e., the narrow Drude peak ($\omega< 0.02 {\rm eV} $)
and the broad incoherent component extending up to $\omega \sim 1 {\rm eV}$
\cite{okimoto}.
The Drude weight is very small, which seems to be consistent with the 
photo-emission spectra.

\noindent
[d] The low temperature resistivity 
$\rho(T)$ can be fitted by
\begin{equation}
\rho(T) = \rho_0 + A T^2,
\end{equation}
where $A$ is a large constant of the 
order of 500$\mu \Omega {\rm cm}/ {\rm K}^2$ \cite{urusibara}, 
again suggesting the strong electron correlation.

\noindent
[e] Contrary to the case of resistivity [d],  the 
coefficient of $T$-linear specific heat is very small 
with $\gamma = 2{\rm mJ/K}$, which violates the Kadowaki-Woods law 
for these compounds \cite{hinetsu}.

Although [e] is difficult to reconcile with [a]-[d], we consider the latter
as the evidence for the strong correlation and orbital
fluctuation even at $T \ll T_c$.
Because the spins are perfectly aligned at $T \ll T_c$, 
the only remaining degrees of freedom are the orbital ones. 
In this section, we study  a
large $d$ model \cite{larged} for the ferromagnetic state 
including both the electron-electron interaction $U$ and the 
Jahn-Teller coupling $g$ \cite{orb}. This model takes care of the 
following two respects, 
(i) including the electron-electron interation, (ii)
including the quantum fluctuations. Especially the latter is 
essential to describe the low temperature Fermi liquid 
state, which is described by the Kondo peak in the  large $d$ limit 
\cite{larged}.
The strong electron-electron interaction with the 
reasonable magnitude of the Jahn-Teller coupling explains both the 
large isotope effect and [a]. Moreover, the features of strong correlation 
[b]-[d] are at least consistent with the large $U$ picture although 
[e] still requires another physics, which we will discuss in the 
next section.
 We start with the Hubbard-Holstein model for the ferromagnetic state. 
\begin{eqnarray}
H&=&-\sum_{i,j, \alpha, \alpha'} 
    t^{\alpha \alpha'}_{ij} c_{i \alpha}^\dagger  c_{j \alpha'}  
   +U \sum_i n_{i \uparrow} n_{i \downarrow} 
 \nonumber \\
  &-& g \sum_i Q_i ( n_{i \uparrow} - n_{i \downarrow} )
 + { 1 \over 2} \sum_i ( {{P_i^2} \over M}  + M \Omega^2 Q_i^2 ),
\end{eqnarray}
where the spin indices $\alpha = \uparrow, \downarrow$ correspond to the 
orbital degrees of freedom as $\uparrow = d_{x^2-y^2}$ and 
$\downarrow = d_{3z^2-r^2}$, and the real spins are assumed to be perfectly 
aligned. We consider only one  Jahn-Teller displacement mode $Q_i$ for each 
site, while there are two modes $Q_{2i},Q_{3i}$ in the real perovskite
structure \cite{kanamori}. 
However this does not change the qualitative features obtained
below. The Jahn-Teller mode is assumed to be an Einstein phonon with
a frequency $\Omega$. The transfer integral 
$t^{\alpha \alpha'}_{ij}$ depends on a pair of  orbitals 
$(\alpha,\alpha')$ and the hopping direction. These dependences lead to 
the various low lying orbital configurations, and 
thus they suppress  the orbital orderings in the ferromagnetic state.
Actually there is no experimental evidence for the orbital orderings,
e.g., the anisotropies of the lattice constants and/or transport properties 
in the ferromagnetic state. Then we assume that no orbital 
ordering occurs down to the zero temperature due to the quantum fluctuations,
and the transfer integral is assumed to be diagonal for simplicity, i.e.,
$t^{\alpha \alpha'}_{ij} = t_{ij} \delta_{\alpha \alpha'}$.
This means that the ground state is a Fermi liquid with two degenerate 
bands at the Fermi energy. In order to describe this Fermi liquid state,
we employ the large-$d$ approach where the model Eq.(3) is mapped to the 
impurity Anderson model with the self-consistent condition \cite{larged}. 
This approach has been applied to the manganese oxides to study
Hund's coupling by Furukawa \cite{furukawa} and to study the 
additional Jahn-Teller coupling  by Millis {\it et al.}  \cite{millis2}.
The action for the effective impurity model is given by
\begin{eqnarray}
S &=& \int^{\beta}_0  d \tau \int_0^\beta  d \tau' 
c^\dagger_{\alpha}(\tau)  G_0^{-1}(\tau-\tau')    c_{\alpha}(\tau') 
 \nonumber \\
&+&  \int_0^\beta  d \tau  \biggl[ U n_{\uparrow}(\tau) n_{\downarrow}(\tau) 
- g Q(\tau) ( n_{ \uparrow}(\tau) - n_{ \downarrow}(\tau) ) \biggr]
+ \int_0^\beta  d \tau  { M \over 2} 
\biggl[ (\partial_\tau Q(\tau) )^2 + \Omega^2 Q(\tau)^2 \biggr],
\end{eqnarray}
where 
$G_0^{-1}(\tau-\tau')$ is the dynamical Weiss field representing the 
influence from the surrounding sites. The self-consistency condition is 
that the on-site Green's function 
$G(i \omega_n) = ( G_0(i \omega_n)^{-1}
- \Sigma(i \omega_n) )^{-1}$  should be equal to the Hilbert 
transform of the density of states $D(\varepsilon)$ as
\begin{equation}
G(i \omega_n) = \int d \varepsilon \,
{ { D(\varepsilon)} \over 
{ i \omega_n + \mu - \Sigma(i \omega_n) - \varepsilon } }.
\end{equation}
We take the Lorentzian density of states  
$D(\varepsilon) = t/(\pi ( \varepsilon^2 + t^2))$ because 
there is no need to solve the self-consistency equation in this case, and
the Weiss field is given by 
\begin{equation}
G^{-1}_0(i \omega_n) =  i \omega_n + \mu + i\, t \,{\rm sign} \,\omega_n. 
\end{equation}
The only unknown quantity is the chemical potential $\mu$,
which is determined by the electron number. 
The determination of chemical potential
requires some numerical calculations, but our discussion below is not
sensitive to the location of the chemical potential. Then we take $\mu$ as the 
parameter of our model, and the problem is now completely reduced  to that of 
a single impurity.
Now we introduce a  Stratonovich-Hubbard (SH) field $\xi(\tau)$ to 
represent the Coulomb interaction $U$. Then the 
action is given by
\begin{eqnarray}
\FL
& &S =\sum_{\omega_n}  
 (i \omega_n + \mu + i \,t \,{\rm sign} \,\omega_n )
c^\dagger_{\alpha} (\omega_n) c_{\alpha}(\omega_n) 
 \nonumber \\
& &+  \int_0^\beta  d \tau  \biggl[ 
{ {\Delta^2} \over U} \xi(\tau)^2 + 
{ M \over 2} [ (\partial_\tau Q(\tau) )^2 + \Omega^2 Q(\tau)^2 ]
+ \zeta(\tau) Q(\tau)
- ( \Delta \xi(\tau) + g  Q(\tau) ) 
( n_{ \uparrow}(\tau) - n_{ \downarrow}(\tau) ) \biggr],
\end{eqnarray}
where 
$\Delta = ( t^2 + \mu^2)/t$ and 
$\zeta(\tau)$ is the test field to measure the phonon correlation function.
At this stage the electron is coupled with the linear combination of the 
SH field $\xi$ and the phonon $Q$ as
$\eta(\tau) \equiv \xi(\tau) + {g \over \Delta}  Q(\tau)$. 
Then the effective action can be derived as
\begin{eqnarray}
S &=& \sum_{\omega_n} \biggl[ 
(i \omega_n + \mu + i t {\rm sign} \omega_n )
c^\dagger_{\alpha} (\omega_n) c_{\alpha}(\omega_n) 
- \Delta \eta(i \omega_n) 
( n_{ \uparrow}(- i \omega_n ) - n_{ \downarrow}(- i \omega_n) ) \biggr] 
\nonumber \\
&+&  \sum_{\omega_n} { 1 \over {2M(\omega_n^2 + {\tilde \Omega}^2 ) }}
\biggl[ 
- \zeta(i \omega_n) \zeta(- i \omega_n)
+ {{2 \Delta^2} \over U} M(\omega_n^2+ \Omega^2) 
\eta(i \omega_n) \eta(- i \omega_n)
\nonumber \\
&+& { {2 g \Delta} \over U }
( \zeta(i \omega_n) \eta(- i \omega_n)
+\eta(i \omega_n) \zeta(- i \omega_n) ) \biggr],
\end{eqnarray}
where ${\tilde \Omega} = \Omega \sqrt{U_{\rm eff.}/U}$.
The effective interaction $U_{\rm eff.}$ is the sum of the 
Coulomb repulsion $U$ and the lattice relaxation energy $E_{\rm LR} = 
g^2/(2M\Omega^2)$ as $U_{\rm eff.} = U + 4E_{\rm LR}$.
The phonon Green's function 
$d(i \omega_n) =  \langle Q(\ i \omega_n) Q(-i \omega_n) \rangle$
is given by 
\begin{equation}
d(i \omega_n) =  
{ 1 \over {M(\omega_n^2 + {\tilde \Omega}^2 ) }}
+ 4 \biggl( { {g \Delta / U}  \over { M(\omega_n^2 + {\tilde \Omega}^2 ) }}
\biggr)^2 \chi_\eta( i \omega_n),
\end{equation}
where 
$\chi_\eta( i \omega_n) = \langle \eta(i \omega_n) \eta(- i \omega_n) \rangle$ 
is the orbital susceptibility of the electronic system.
The first term is the usual phonon Green's  function
with the renormalized frequency ${\tilde \Omega}$, which is higher 
than the bare $\Omega$. This 
${\tilde \Omega}$ does not depend on the electron response function
and hence temperature independent. 
When $U \ll E_{\rm LR}$, ${\tilde \Omega}$ is much larger than $\Omega$
and the first term  of Eq.(9) is irrelevant in the phonon frequency region 
$\omega \sim \Omega \sim 0.1 {\rm eV}$.
In the opposite limit $U \gg  E_{\rm LR}$, the 
renormalization of the phonon frequency is small. 
The second term depends on the 
temperature (Kohn anomaly) and shows the characteristic lineshape with the
broadening.  
From the above  consideration together with the experimental fact that 
no temperture dependence of the phonon frequency has been observed, 
we conclude that the Coulomb interaction $U$ is larger than the lattice 
relaxation energy $E_{\rm LR}$.
\par
Now let us study the electronic system in detail.
Following the analysis in section III C of ref.\cite{hamann},
the quantum fluctuation between two orbital configurations gives rise
to the orbital Kondo effect with the Kondo temperature $T_K$.
We obtain an  estimate for the Kondo temperature as
\begin{equation}
T_K \sim t  \exp\biggl[ - {{RU_{\rm eff.}} \over { \Delta}} \biggr].
\end{equation}
where 
\begin{eqnarray}
R &=& 2x \sqrt{ 1 + 4 y^{-1}},\quad {\rm for}\;\; y >> 1/\max(x,x^2) \\
R& =& \sqrt{ 1 + { 2 \over 3} x^2(4+y) }, \quad {\rm for} \;\; y << 1/\max(x,x^2).
\end{eqnarray}
Here we have introduced the dimensionless
variables  $x = E_{\rm LR}/\Omega$ and $y = U/E_{\rm LR}$. 
The main conclusion here is that when $x>1$, i.e., $E_{\rm LR} > \Omega$,
the reduction factor $R$ is proportional to $x$ even in the limit
 $U \gg E_{\rm LR}$.
This is because the overlap of the phonon wavefunction enters the 
tunneling matrix element even when the Coulomb interaction is dominating.

This Kondo temperature $T_K$ gives the effective bandwidth $B$ in
the large-$d$ model, and the effective mass enhancement is 
estimated as $m^*/m = t/T_K \sim
\exp\biggl[ {{RU_{\rm eff.}} \over { \Delta}} \biggr]$.
This strong mass enhancement manifests itself in the physical quantities 
as follows.
The orbital susceptibility $\chi_{\eta}(i \omega_n \to \omega + i \delta)$
has a peak at $\omega=0$ with the peak height of the order of 
$\eta_0^2/T_K$ and with the width of the order of $T_K$.
The specific heat  coefficient $\gamma$ should be proportional to 
$B^{-1}$, while the coefficient $A$ of $T^2$ in Eq.(2) scales as
$A \propto B^{-2}$. 
The ferromagnetic transition temperature $T_c$ is expected 
to be proportional to $B$, and hence we expect a large  
isotope effect when $R>1$.
In other words, the large isotope effect does not rule out the large 
Coulomb interaction $U$.
The edge of the photoemission spectrum at the Fermi energy and the 
Drude weight are reduced by the factor $m/m^*$.
As mentioned above, many experiments except for those of  
specific heat seem to suggest that the manganese oxides in their
ferromagnetic state belong to the strongly correlated region.
Then it is natural that the large isotope effect is observed.
The band width is rather difficult to pin down accurately, but a
rough estimate from $A$ in eq.(2) is 
$B \sim 0.1 {\rm eV}\,$, which is comparable to $\Omega$.
This suggests that the suppression is not extremely large. Another 
experimental fact is that the temperature independent Jahn-Teller 
phonon is observed at the frequency near the original one \cite{yamamoto}, 
which means that $U$ is larger or at least of the order of $E_{\rm LR}$.
Then we conclude that the relative magnitude of the 
energy scales is $\Omega<E_{\rm LR}<U$ for the ferromagnetic phase in the 
manganese oxides.
The small specific heat coefficient necessitates 
the physical mechanisms which have not been included 
in our model, which will be discussed in the  next section.  

Summarizing this section,  we have studied a model of manganese 
oxides in the ferromagnetic state taking into account both the 
Coulomb repulsion $U$ and the Jahn-Teller 
interaction $E_{\rm LR}$. These two interactions collaborate to induce the 
local orbital moments.
In the strong coupling case, i.e., $U_{\rm eff.} =U+4E_{\rm LR}>> t$
($t:$ transfer integral), the overlap of the phonon wavefunctions
($\sim e^{-E_{\rm LR}/\Omega}$) enters the tunneling amplitude between the 
two minima for the orbital moment.  
The phonon spectral function consists of two parts, i.e., 
the sharp peak at the renormalized frequency ${\tilde \Omega}
= \Omega \sqrt{ U_{\rm eff.}/U}$ and the broad peak with the 
width of the order of the band width.
These results can reconcile the
large isotope effect on $T_c$ and  the apparent temperature independent 
phonon spectrum assuming $U > E_{\rm LR}>\Omega$.
\section{Conclusions}

We have studied the vital role of the orbitls in manganese
oxides. The most important ingredient is the strong Coulombic interaction,
which induces the orbital polarization. This polarization controls the 
dimensionality of the electronic system when ordered.
In the isotropic ferromagnetic state, large orbital 
fluctuation is expected which  
gives rise to the quantum liquid state below the orbital Kondo
temperature $T_K$.
>From this viewpoint, the double exchange interaction is switched on below
the orbital Kondo temperature $T_K$, and the ferromagnetic transition 
temperature $T_c$ is roughly identified with $T_K$.
This scenario might explain the photoemission experiment
which shows that the discontinuity at the Fermi level grows only below 
$T_c$.
However there still remain several puzzles as follows. 
(i) the small specific heat, (ii) violation of the Kadowaki-Woods relation,
(iii) absence of the quasi-particle peak in the angle-resolved
photoemission spectra in double layered manganese oxides,
(iv) the residual resistivity larger than the Mott limit.
One possible explanation for these is the (dynamical) phase separation.
The phase separation occurs when the total energy $E(x)$ as a function
of the hole concentration $x$ has more than two minima and 
it is possible to draw a common tangential line. 
This is found to occur for several models. Here we want to point out that
when the bandwidth strongly depends on the orbital configuration,
the $x$-dependence of it naturally gives rise to the phase separation.
In reality the long range Coulomb interaction will suppress the 
phase separation, but which still remain as a fluctuation, i.e.,
dynamical one.
Suppose phase separation occurs, and assume that $r$ is the
metallic fraction while $1-r$ is the insulating fraction.
Now we relate the observed physical quantities to the 
metallic ones and the fraction $r$.
The $T$-linear specific heat $\gamma_{\rm obs}T$
is reduced by the factor $r$, and hence 
$\gamma_{\rm obs} = r \gamma_{\rm metal}$.
Concerning the resistivity, the percolating path will lower the 
capacitance energy when the voltage is applied, and 
the observed resistivity is enhanced by the factor $1/r$,
i.e., $\rho_{\rm obs}(T) = \rho_{\rm metal}(T)/r$.
Therefore the Kadowaki-Woods ratio is related to 
\begin{equation}
\biggl({ A \over {\gamma^2} } \biggr)_{\rm obs} =
r^{-3}\biggl({ A \over {\gamma^2} } \biggr)_{\rm metal}. 
\end{equation}
The observed ratio is $10^2$ times larger than the universal ratio, 
and if we attribute this discrepancy to the factor $r$,
we obtain $r \cong 0.2$.
This also suggests $\gamma_{\rm metal} \cong 14 mJ/mol$, which means 
that the mass enhancement is $\sim 10$.
Furthermore the Drude wieght is reduced by $r$ in addition to the mass
enhancement, and is expected to be about $1/50$ of the noninteracting
value in agreement with experiment.
The residual resistivity is enhanced by $1/r$  and can be larger than
the Mott limit. Roughly speaking, the Hall constant, 
on the other hand, is not modified by the phase separation. 
The extremely large noise might be also related to the 
dynamical fluctuation of the phase separation.
Thus the phase separation seems to explain many puzzles in the 
manganese oxides. However the detailed studies including the 
time/length scale of the fluctuation are left for future 
investigations.

The authours would like to thank  
Y.Tokura, A. Millis, J. Ye, G. Kotliar, Q. Si, 
H. Kuwahara, S. Maekawa, H.Fukuyama, K. Miyake, 
T.Okuda, K.Yamamoto and Y.Endoh for valuable discussions. 
This work was supported by COE and Priority Areas Grants from 
the Ministry of Education, Science, Culture and Sports of Japan.
Part of this work has been done in APCTP workshop and
Aspen Center for Physics, and we acknowledge their
hospitalities.
%

%
\vfill 
\eject
\noindent
Figure captions
\par
\noindent
\\
Figure 1. 
 Mean field phase diagram as a function of the carrier 
concentration ($x$) and the antiferromagnetic interaction 
between $t_{2g}$ spins ($J_S$). 
The energy parameters are chosen to be $\tilde \alpha=8.1$ 
and $\tilde \beta =6.67$. 
Schematic orbital structure in the each phase 
is also shown. 
Dotted line ($J_S$ =0.009) 
well reproduces the change of the spin structure 
experimentally observed.
\par
\noindent
\par \noindent

\begin{references}
%
\bibitem{Zener}
 C. Zener, Phys. Rev. {\bf 82}, 403 (1951).
%
\bibitem{Anderson}
 P. W. Anderson, and H. Hasegawa, Phys. Rev. {\bf 100}, 675 (1955).
%
\bibitem{deGenne}
 P. G. de Gennes, Phys. Rev. {\bf 118}, 141 (1960).
%
\bibitem{ishi96}
S. Ishihara, J. Inoue, and S. Maekawa, 
Physica C {\bf 263}, 130 (1996), and Phys. Rev. B {\bf 55}, 8280 (1997). 
%
\bibitem{ishi97}
 S. Ishihara, M. Yamanaka, N. Nagaosa, Phys. Rev. B {\bf 56}, 686 (1997). 
%
\bibitem{maezono1}
 R. Maezono, S. Ishihara and N. Nagaosa,
Phys. Rev. B {\bf 57}, R21822 (1998). 
%
\bibitem{maezono2}
 R. Maezono, S. Ishihara and N. Nagaosa,
to be published in Phys. Rev. B {\bf 58}, November (1998).
%
\bibitem{Woll}
 E. O. Wollan, and W. C. Koehler, Phys. Rev. {\bf 100}, 545 (1955).
%
\bibitem{Kawano2}
 H. Kawano, R. Kajimoto, H. Yoshizawa, Y. Tomioka, H. Kuwahara
 , and Y. Tokura, Phys. Rev. Lett. {\bf 78}, 4253 (1997).
%
\bibitem{Kuwahara1}
H. Kuwahara, T. Okuda, Y. Tomioka, T. Kimura, A. Asamitsu and 
Y. Tokura, Mat. Res. Soc. Sym. Proc. {\bf 494}, 83 (1998).
%
\bibitem{PrSr}
Y. Tomioka (unpublished).
%
\bibitem{LaSr}
 Y. Moritomo, T. Akimoto, A. Nakamura,
 K. Ohoyama, and  M. Ohashi,
 Phys. Rev. B {\bf 58}, 5544 (1998).
%
\bibitem{endoh}
M.C.Martin, G.Shirane, Y.Endoh, K.Hirota, Y.Moritomo, Y.Tokura,               
Phys. Rev. {\bf 53}, 14285 (1996).
%
\bibitem{yamamoto}
K.Yamamoto and Y.Tokura, private communications.
\bibitem{sarma}
D.D.Sarma {\it et al.}, Phys. Rev. B{\bf 53}, 6874 (1996).         
\bibitem{okimoto}
Y.Okimoto {\it et al.}, Phys. Rev. Lett. {\bf 57}, 109 (1995).
\bibitem{urusibara}
A.Urushibara {\it et al.}, Phys. Rev. B{\bf 51}, 11103 (1995).
\bibitem{hinetsu}
B.Woodfield, M.L.Wilson, and J.M.Byers,
Phys. Rev. Lett. {\bf 78}, 3201 (1997); T.Okuda and Y.Tokura, 
Private communications.
\bibitem{larged}
For a review see A.Georges {\it et al.},
Rev. Mod. Phys. {\bf 68}, 13 (1996).
%
\bibitem{orb}
N. Nagaosa, S. Murakami, and H.C.Lee,
Phys. Rev. B {\bf 57}, R6767 (1998). 
%
\bibitem{kanamori}
J.Kanamori, J. Appl. Phys.{\bf 31}, 14S (1960).
%
\bibitem{furukawa}
N.Furukawa, J. Phys. Soc. Jpn {\bf 63}, 3214 (1994).
%
\bibitem{millis2}
A.J.Millis, R.Mueller, and B.I.Shraimain, Phys. Rev. B{\bf 74}, 
5389; 5405 (1996).
%
\bibitem{hamann}
D.R.Hamann, Phys. Rev. {\bf 2}, 1373 (1970).
\end{references}
\end{document}